\documentclass[preprint,showpacs,preprintnumbers,amsmath,amssymb,superscriptaddress]{revtex4}
\usepackage{graphicx}
\usepackage{amsmath}
\usepackage{amssymb}
\usepackage{epsfig}
\usepackage{amsthm}
\usepackage{color}
\usepackage{animate}
\def\be{\begin{equation}}
\def\ee{\end{equation}}
\def\arr{\begin{array}{rll}}
\def\ea{\end{array}}
\def\bea{\begin{eqnarray}}
\def\eea{\end{eqnarray}}
\begin{document}

\title{Extended symmetries in geometrical optics}

\author{Zhyrair Gevorkian}
\affiliation{Yerevan Physics Institute, Alikhanian Brothers St. 2, 0036, Yerevan, Armenia}
\affiliation{Institute of Radiophysics and Electronics, Ashtarak-2,0203, Armenia }
\author{Mher Davtyan}
\affiliation{Institute of Radiophysics and Electronics, Ashtarak-2,0203, Armenia }
\author{Armen Nersessian}
\affiliation{Yerevan Physics Institute, Alikhanian Brothers St. 2, 0036, Yerevan, Armenia}
\affiliation{Yerevan State University, 1 Alex Manoogian St., Yerevan, 0025, Armenia}
\begin{abstract}

We examine additional symmetries of specific refraction index profiles that are used in the well-known phenomena of perfect imaging and cloaking.In the considered cases, the translation generator and the angular momentum are conserved. We express the ray trajectory parameters through the integrals of motion and observe the existence of a photon state with maximal angular momentum, which can be used as an optical resonator. Application in plasmons and the role of polarization are discussed, and the Spin Hall effect in an extended symmetry profile is predicted.

\end{abstract}
\maketitle

\section{Introduction}
Cloaking phenomena have attracted a great deal of interest since Pendry \cite{Pendry06} and Leonhardt \cite{Leonhardt06}, who assumed that an object coated by certain inhomogeneous shell becomes invisible to electromagnetic waves. Different mechanisms of cloaking have been suggested since that time, among which are anisotropic metamaterial shells \cite{Pendry06}, conformal mapping in two-dimensional systems \cite{Leonhardt06}, complementary media \cite{lai09}, etc (for a review see \cite{gbur13} and references therein).
Transformation optics \cite{Pendry06} is the most frequently used approach. In this approach, the dielectric permittivity and magnetic permeability tensors are specific coordinate-dependent functions. However, this approach is difficult to implement in the optical field, since it is problematic to find metamaterials with the necessary magnetic properties \cite{zhou05}. In the cases when the photon wavelength is much smaller than the characteristic size of inhomogeneity, geometrical optics approximation is justified \cite{Sun08}-\cite{choi14}. In the conformal mapping method of cloaking when geometrical optics approximation is used,  closed ray trajectories are of significant importance \cite{Leonhardt06}. Note that these trajectories determine the size and shape of the cloaked area \cite{Leonhardt06}.

In the present paper, we thoroughly consider the closed ray trajectories within geometrical optics approximation with the emphasis on the symmetries of the refraction index profile. Namely, we consider the refraction indices which yield  the optical metrics  coinciding with those of three-dimensional sphere (for positive $\kappa$) and two-sheet hyperboloid/ pseudosphere (for negative $\kappa$),
 \begin{equation}
n({\bf r})=\frac{n_0}{|1+\kappa{\bf r}^2|},\qquad \kappa=\pm\frac{1}{4r^2_0}.
\label{ns}
\end{equation}
While the generic homogeneous spaces have $so(3)$ symmetry algebra generated by conserved angular momentum, three-dimensional sphere and pseudosphere have symmetry algebras  $so(4)$ and $so(3.1)$, respectively. The reason for that is the existence of three additional conserved quantities - ``translation generators".
The two-dimensional counterpart of such refraction index with a positive sign along with other profiles corresponding to the two-dimensional Hamiltonian systems with closed trajectories \cite{Leonhardt06}was already used to describe cloaking phenomena in $2d$ via conformal mapping. This profile is well-known  in optics  as  ``Maxwells fish eye" \cite{Maxwell},\cite{bowolf} and was  debated as a possible tool for perfect imaging \cite{Leonhardt09,Philbin10,Blaikie10}.

In our opinion, the importance of the present consideration is the direct use of extended symmetry of the sphere and pseudosphere. Seemingly, it might be extended to other three-dimensional profiles corresponding to the Hamiltonian systems with close trajectories, such as three-dimensional oscillator and Coulomb potential and their deformations with Calogero-like potentials, as well as their generalizations to three-dimensional spheres and pseudospheres  \cite{CalCoul}. The consideration of the three-dimensional (two-sheet) hyperboloid which can be used for the study of plasmon perfect imaging, cloaking is another important point in the given study. Here we have mentioned cloaking and perfect imaging as possible applications of closed ray trajectories. However, there can be other applications too.

In this particular paper,an we mostly neglect the light polarization, whose interaction with the inhomogeneity of dielectric permittivity leads to many well-known effects, e.g. the optical Hall effect \cite{Mur04, bnature, bliokh}, capsize of polarization and straightening of light in dilute photonic crystals \cite{ghgc17,ggc19}, etc. We only briefly discuss its influence on the trajectories postponing the detailed consideration for the future study.

\section{Initial Relations}
It is well-known  that minimal action principle came \textit{to} physics from the
geometric optics. Initially, it was invented for the description of
the propagation of light in media by Fermat, and is presently known as the Fermat principle,
\begin{equation}
 {\cal S}_{Fermat} =\frac{1}{\lambdabar_{0}}\int d{s},\qquad ds=n({\bf r}) |d{\bf r}/d\tau |d\tau
\label{gactions2}\end{equation}
where  $n({\bf r})$ is the
refraction index, and $\lambdabar_0$ is wavelength in vacuum which defines the length of the light trajectory under the
assumption that the helicity of light is neglected. This  action could be interpreted as the action of the system on the
three-dimensional curved space equipped with the ``optical metrics" (or Fermat metrics)
of Euclidean signature
 \begin{equation}
  ds^2= g_{AB} dx^A dx^B,\qquad
g_{AB}=n^2({\bf r})\delta_{AB},\qquad A,B=1,2,3
\label{om}
\end{equation}
In the special case when the
refraction index has a form \eqref{ns},
the Fermat metrics  describes  three-dimensional sphere (for  $\kappa>0$), or two-sheet hyperboloid (for  $\kappa<0$) \textit{with} radius $r_0$.
In this case, in addition to rotational,  $so(3)$, symmetry  which yields the conserving angular momentum, the system has $so(4)$/$so(3.1)$ symmetry which provides the  three-dimensional sphere/hyperboloid, by  three additional symmetries conserved quantities .
In the case of positive $\kappa$, the refraction index is a decreasing function of $r$, while for the negative $\kappa$, it has an increasing part (see Fig.1).
\begin{figure}
 \begin{center}
\includegraphics[width=16.0cm]{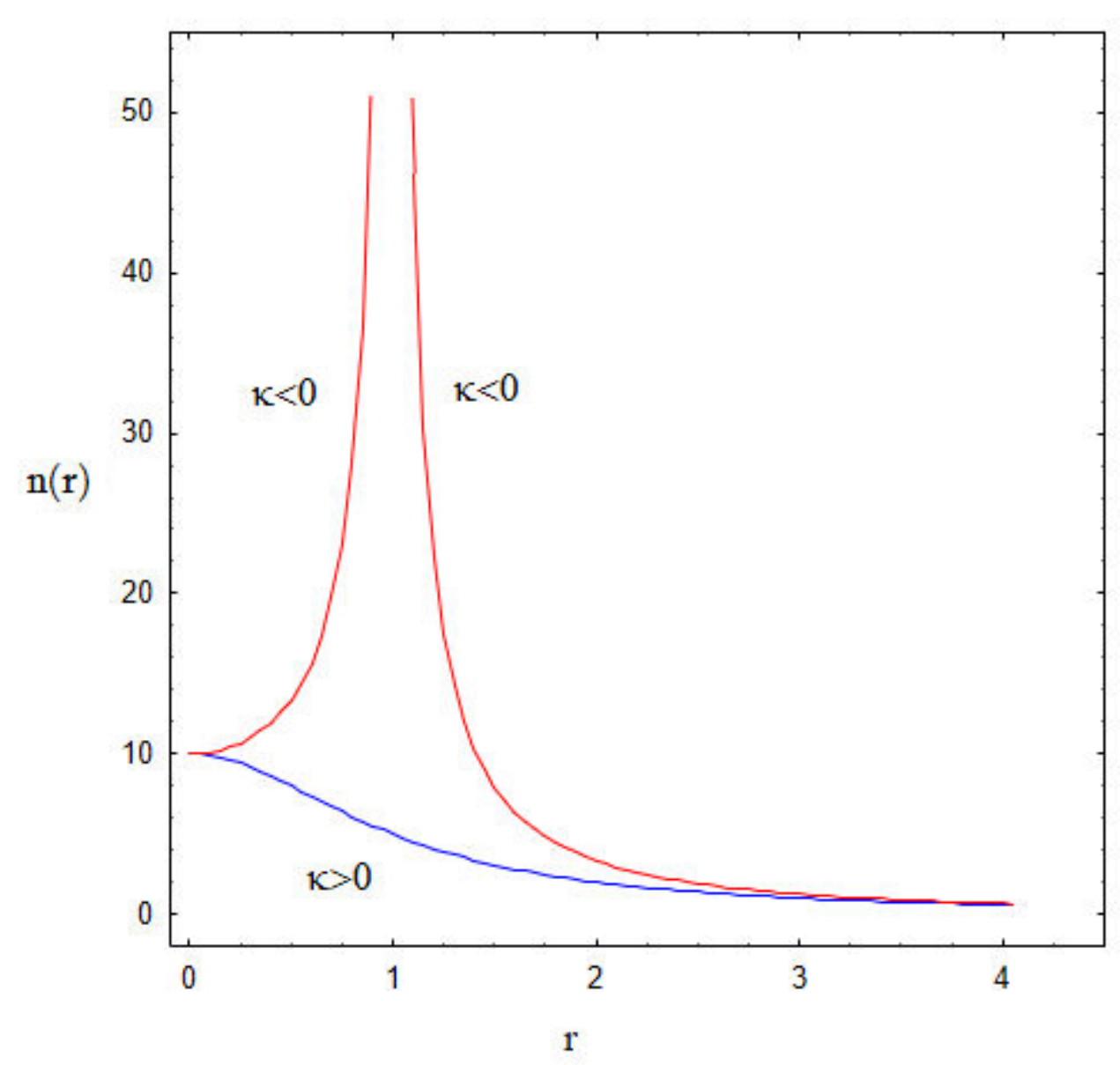}
\caption{Refraction index profiles}
\label{fig.1}
\end{center}
\end{figure}
In this paper, we show that due to the extended symmetry mentioned above, these very different profiles lead to closed ray trajectories.
As noted in the Introduction,  the case of the positive sign was already used to describe cloaking phenomena in $2d$ via conformal mapping \cite{Leonhardt06}.

Let us give the Hamiltonian formulation of the system defined by the action \eqref{gactions2}.
Due to its reparametrization-invariance, the Hamiltonian constructed by the standard Legendre transformation is identically zero. However, following Dirac's theory \cite{dirac}, the constraint between  momenta  and coordinates appears as follows
 \begin{equation}
  \Phi\equiv g^{ij}(x)p_ip_j-\lambdabar^{-2}_0 =0, \quad{\rm with}\quad g^{ij}p_ip_j=\frac{{\bf p}^2}{n^2({\bf r})},\qquad {\rm with}\quad g^{ij}g_{jk}=\delta^i_k.
 \label{constraint}
 \end{equation}
 Hence, the Hamiltonian system corresponding to the action \eqref{gactions2},
is defined by the canonical Poisson brackets
 \begin{equation}
  \{f,g\}=\frac{\partial f}{\partial x_\alpha}\frac{\partial g}{\partial p_\alpha}-\frac{\partial f}{\partial p_\alpha}\frac{\partial g}{\partial x_\alpha}
 \label{pb0}\end{equation}
and by the  Hamiltonian
\begin{equation}
\mathcal{H}_0=\alpha({\bf p},{\bf r})\Phi =\alpha({\bf p},{\bf r})\left( \frac{{\bf p}^2}{n^2({\bf r})}-\lambdabar^{-2}_0 \right)\approx 0.
\label{h0}
\end{equation}
Here $\alpha$ is the Lagrangian multiplier, which could be an arbitrary function of coordinates and momenta.
When we write down the Hamiltonian equations of motion, the notation ``weak zero" ($\mathcal{H}_{0}\approx 0$) indicates that we should take into account the constraint \eqref{constraint} only after differentiation,
\be
{\dot f}({\bf r}, {\bf p})=\{f, \mathcal{H}_0\}=\{f,\alpha\}\Phi+\alpha \{f,\Phi\}\approx\alpha\{f,\Phi\}.\ee
The  arbitrariness in the choice of the function $\alpha$ reflects the reparametrization-invariance of the action \eqref{gactions2}.
Suppose, for the description of the  equations of motion in terms of arc-length of the original Euclidian space  one should choose (see \cite{bliokh})
\begin{equation}
\alpha^{-1} =\frac{|{\bf p}|+ \lambdabar^{-1}_0 n({\bf r})}{n^2({\bf r})},\qquad \Rightarrow\quad \mathcal{H}_0=|{\bf p}|- \lambdabar^{-1}_0n({\bf r})
\label{alpha}
\end{equation}
With this choice, the Hamiltonian equations of motion take the conventional form \cite{ko}
 \begin{equation}
{\bf \dot{p}}=\lambdabar^{-1}_0{\bf \nabla} n(\bf r),\qquad
{\bf \dot{r}}={\bf p}/{|{\bf p}|}.
\label{hameq}
\end{equation}
These equations describe the motion of a wave package with center coordinate ${\bf r}$ and momentum ${\bf p}$ in a curved space.
However, for preserving the similarity with classical mechanics we will deal  with the generic formulation \eqref{h0}, i.e. we will not fix the parametrization of light rays.

We are interested in the integrals of motion, i.e. physical quantities that are conserved along the ray trajectory.
For the isotropic media, when  the refraction index is a spherical symmetric function, $n({\bf r})=n(|{\bf r}|)$, the angular momentum of the system is conserved
\begin{equation}
\mathbf{L}=\mathbf{r}\times \mathbf{p}.
\label{moment}
\end{equation}
When the refraction index \textit{has} the form \eqref{ns},  the Fermat   metrics coincide with those of three-dimensional sphere (for $\kappa >0$) or two-sheet hyperboloid (for $\kappa <0$) written in conformal flat coordinates (we ignore here $n_0$ factor),
 \begin{equation}
\frac{d{\bf r}\cdot d{\bf r}}{(1\pm \frac{{\bf r}^2}{4r^2_0})^2}=\left(d{\bf y} \cdot d{\bf y} \pm dy^2_4\right)\mid_{ y^2_4\pm {\bf y}^2=r^2_0} \; ,
 \end{equation}
where
 \begin{equation}
 y_4=-r_0\frac{1-\kappa {\bf r}^2}{1+\kappa {\bf r}^2},\quad {\bf y}=\frac{{\bf r}}{1+\kappa {\bf r}^2},\quad {\rm with }\quad \kappa =\pm\frac{1}{4r^2_0}
 \label{amc}
 \end{equation}
 From Eq.\eqref{amc} we get
 \begin{equation}
 {\bf r}=2r_0\frac{{\bf y}}{r_0+y_4}\quad \Rightarrow\quad {\bf r}^2=4r^2_0\frac{r_0-y_4}{r_0+y_4}
 \label{rest}
 \end{equation}
This is just stereographic projection of the (pseudo)sphere on the ``$\mathbf{r}$-space"  which touches it at the pole $y_4=-r_0$.
The upper and lower hemispheres are projected to the inside and outside of the three-dimensional ball with radius $2r_0$, respectively. The "equatorial sphere" $y_4=0$  is projected to the boundary of that ball,  which is a two-dimensional sphere in the ``$\mathbf{r}$-space". Due to  the $so(4)/so(3.1)$ symmetry of three-dimensional sphere/hyperboloid, in addition to  $so(3)$ algebra generators \eqref{moment},  the system  possesses  three more  conserving quantities
\begin{equation}
\mathbf{T}=(1-\kappa \mathbf{r}^2)\mathbf{p}+2\kappa(\mathbf{pr})\mathbf{r}\quad :\quad \{\mathbf{T},\mathcal{H}_0\}=0.
\label{T0}
\end{equation}
The Hamiltonian becomes  Casimir of   $so(4)/so(3.1)$ algebra(s)
\begin{equation}
\frac{{\bf p}^2}{(1+\kappa r^2)^2}={\bf T}^2+4\kappa{\bf L}^2\quad\Rightarrow \quad  {\bf T}^2+4\kappa{\bf L}^2=\frac{n^2_0}{\lambdabar^{2}_0}.
\label{cas}
\end{equation}
Notice also, that $\mathbf{T}$ is perpendicular to $\mathbf{L}$:  $\mathbf{T}\bot \mathbf{L}$.
\section{Trajectories}
Extended symmetry allows to obtain ray trajectories without solving the equations of motion \eqref{hameq}. Namely, the vector product of   ${\bf T}$ and  ${\bf r}$  immediately yields  the expression of trajectories,
\begin{equation}
\mathbf{L}=\frac{\mathbf{T}\times\mathbf{r}}{1-\kappa \mathbf{r}^2}\quad \Rightarrow\quad     \vert\mathbf{r}-\mathbf{a}_0\vert^2=R^2_0,\qquad \mathbf{a}_0\equiv\frac{\mathbf{T}\times{\mathbf L}}{2\kappa L^2}, \quad R_0=\frac{n_0}{2|\kappa| L\lambdabar_0}.
\label{trajec}
\end{equation}
\textit{Ray} trajectories are the circles with center ${\bf a_0}$ and radius $R_0$ \cite{bowolf}. Note  that  the radius of  circle is independent of the sign of $\kappa$, whereas the coordinates of the center ${\bf a_0}$ depend on the sign of $\kappa$. From \eqref{trajec}, it is easy to find that
  \begin{equation}
  |a_0|=\sqrt{R_0^2-1/\kappa }
  \label{centcor}
  \end{equation}
Using expressions \eqref{centcor} and \eqref{trajec} one can
  draw the ray trajectories Fig.2.
  \begin{figure}
   \begin{center}
\includegraphics[width=16.0cm]{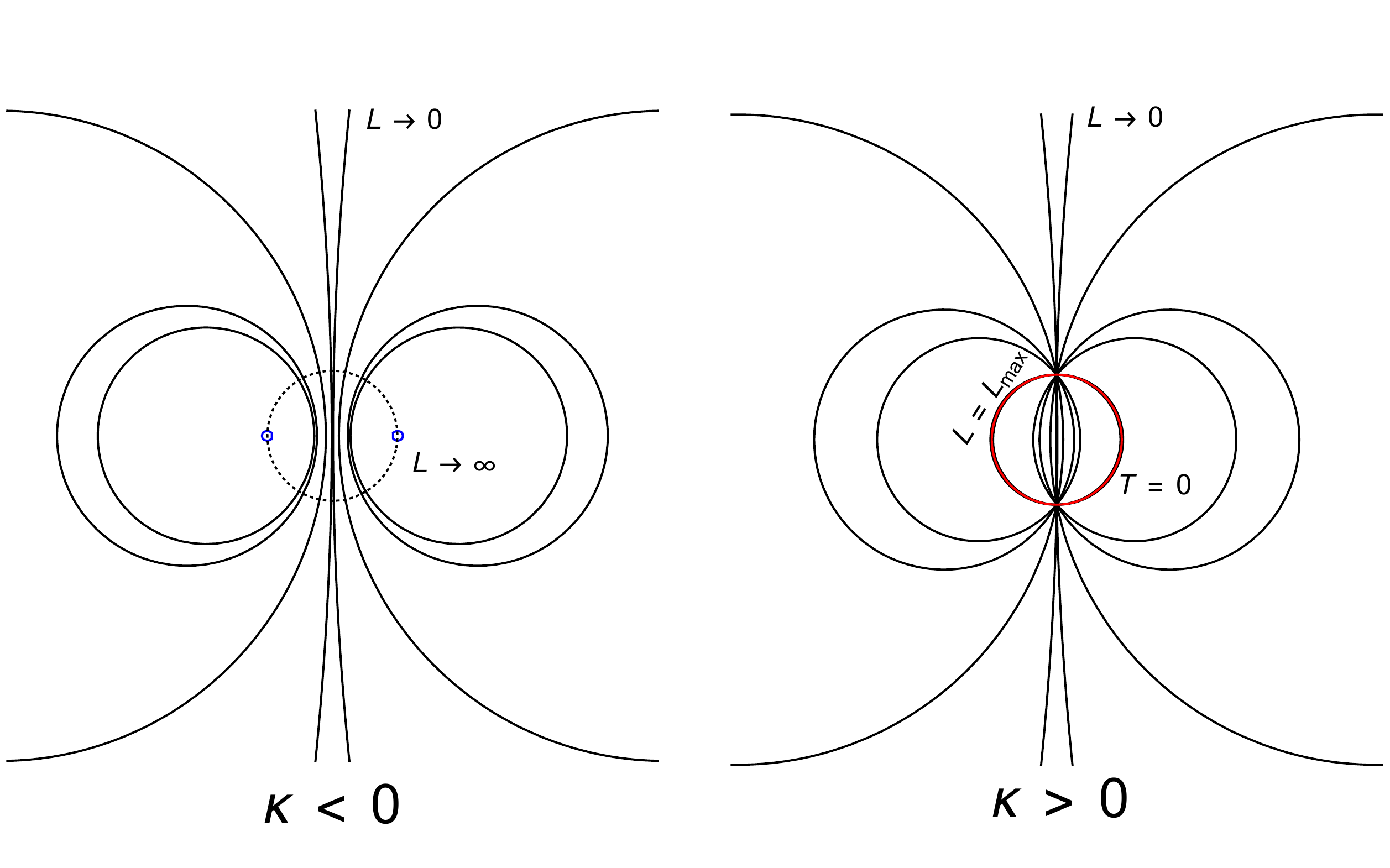}
\caption{Ray trajectories for different values of ${\bf T},{\bf L},\kappa$. Dashed and red circles with radius $2r_0$ are given by the refraction profile \eqref{ns}. The red circle is the photon trajectory with maximal angular momentum of the photon in $\kappa>0$ case. In $\kappa>0$ case, symmetric trajectories with $\pm a_0$ intersect while in $\kappa<0$ case they do not. The straight lines are trajectories without angular momentum $L=0$. The small blue circles are trajectories in $L\to\infty$ limit  provided that $\kappa<0$.}
\label{fig.2}
\end{center}
\end{figure}
For the photons with zero angular momentum $L=0$,  the radius of the circle goes to infinity $R_0\to\infty$, i.e. we get a straight line.
In contrast with common approaches (see, e.g.  \cite{bowolf}), we express the equations of ray characteristics  \eqref{trajec} through the integrals of motion, which allows us to consider different physical situations.

Now, let  us consider the  cases when either ${\bf T}$ or ${\bf L}$ become  zero (they  cannot  be  equal to zero simultaneously due to \eqref{cas}).
 If $\kappa<0$ from Eq. \eqref{cas} it follows that the only possibility is $L=0$ , $T\neq 0$, with $T$ taking minimal value
 $T_{min}=n_0/\lambdabar_0$.
Using  \eqref{centcor},\eqref{cas} one can see that  for $\kappa<0$  one always has $|a_0|>2r_0$ and for $\kappa>0$ one has  $R_0>2r_0$. So, in both cases there are not any ray closed trajectories inside the  area of radius $2r_0$, see Fig.2. In $\kappa>0$ case, photons with angular momentum $L>L_{max}$ will not form closed trajectories and correspondingly can not ensure perfect imaging and cloaking. This restriction is absent in $\kappa<0$ case. In the conformal mapping scheme cloaking area is determined by the outer part of closed trajectories \cite{Leonhardt06}. Therefore from \eqref{trajec}, it follows that cloaking area will disappear for small $L$.

For $\kappa >0$ , $T=0$, $L\neq 0$ angular mommentum  $L$ acquires the maximal  value $L_{max}=n_0r_0/\lambdabar_0$ \eqref{cas}.
As it follows from \eqref{T0},   when $T=0$, $r^2=4r_0^2$ and ${\bf p}\bot {\bf r}$ the  photon trajectory becomes a circle with the radius $R_0=2r_0$ and with center at $\mathbf{a}_0=0$ (see Fig.2).
In this state, the angular momentum of photon  can be very large $n_0r_0/\lambdabar_0\gg 1$. These states are interesting for quantum information purposes \cite{zeil18}. Besides that, this state can be used as an optical resonator that is an accumulator of energy.
To be convinced let us determine electric and magnetic fields on this trajectory. In geometrical optics one can use the following expansions \cite{bowolf}
\begin{equation}
{\bf E}=e^{\imath \frac{\Psi}{\lambdabar_0}}\sum_{m\geq 0}(-\imath \lambdabar_0)^m{\bf e}_m\qquad
{\bf H}=e^{\imath \frac{\Psi}{\lambdabar_0}}\sum_{m\geq 0}(-\imath \lambdabar_0)^m {\bf h}_m,
 \label{exp}
 \end{equation}
 where  ${\bf e_m,h_m}$ are functions of coordinates which  can be found by substituting expressions of $\bf E$ and $\bf H$ from \eqref{exp} into Maxwell equations. Geometrical optics approximation corresponds to leading terms of expansion \eqref{exp}. The equation for eikonal $\Psi$ has the form
\begin{equation}
n({\bf r})\frac{d {\bf r}}{ds} ={\bf \nabla}\Psi
\label{eikeq}
\end{equation}
Substituting \eqref{hameq} into \eqref{eikeq} and using \eqref{h0} we then  find
\begin{equation}
\Psi=\lambdabar_0{\bf pr}
\label{eik}
\end{equation}
As mentioned above, for this trajectory $T=0$, $L=L_{max}=n_0r_0k_0$, ${\bf p}\bot {\bf r}$ and ${\bf p}{\bf r}\equiv 0$. Therefore on this trajectory, as it follows from \eqref{exp}, the phase (eikonal) remains constant during the round trip of the photon. Hence, in this state, photon constructively interferes with itself, and the energy is being accumulated (see also \cite{svelto10,turks14}).

Note that the basic ray trajectory ($T=0$, $L=L_{max}$) with large angular momentum is similar to whispering gallery modes that originate as an eigenstate of dielectric sphere (see \cite{oraevsky02}).



\subsubsection*{Plasmon }

In the $\kappa<0$ case,  the refraction index diverges at the point  $r = 2r_0$. Such a situation can be realized,
for example, on the metal surfaces near the plasmon resonance frequencies.
Indeed, it is  well known that  dispersion equation of plasmon on the interface of a metal with dielectric constant $\varepsilon(\omega)<0$
and with dielectric permittivity $\varepsilon_d$ has the form
\begin{equation}
k_p=\frac{\omega}{c}\sqrt{\frac{\varepsilon(\omega)}{\varepsilon(\omega)+\varepsilon_d}},
\label{plasmon}
\end{equation}
with $\varepsilon(\omega)\approx -\varepsilon_d$ near the plasmon resonance.

Suppose that the dielectric material has an inhomogeneous profile $\varepsilon_d\equiv \varepsilon_d(r)$.  From the expression above, it can be presumed that
a plasmon moves in a $2d$ medium with refraction profile
\begin{equation}
n(r)=\sqrt{\frac{\varepsilon(\omega)}{\varepsilon(\omega)+\varepsilon_d(r)}}.
\label{refract}
\end{equation}
Suppose that  $\mathbf{r}_p$ is a resonance point,  $\varepsilon(\omega)= -\varepsilon_d(r_p)$.
Expanding $\varepsilon_d(r)$ around $r_p$ and assuming that $\varepsilon'(r_p)=0$, we get
\begin{equation}
n(r)\approx \frac{1}{|r-r_p|}\sqrt{\frac{2\varepsilon(\omega)}{\varepsilon_d''(r_p)}}.
\label{appref}
\end{equation}
If we choose $n_0$ and $r_0$ such that
\begin{equation}
n_0r_0=\sqrt{\frac{2\varepsilon(\omega)}{\varepsilon_d''(r_p)}}, \qquad \kappa<0,\quad r_0=\frac{r_p}{2},
\label{compref}
\end{equation}
near  the  resonance  point, $n(r)$ will obtain the form \eqref{ns}.
So, it is possible to choose indexes $\varepsilon_p (r)$ such that
near plasmon resonance point, closed trajectories and therefore cloaking phenomenon via conformal mapping can be realized (see also \cite{plasmon}).

\section{Generalizations}
In the previous section we  related closed trajectories  with the free particles  on the sphere and two-sheet hyperboloid, which are the simplest three-dimensional maximally superintegrable systems (the $N$-dimensional dynamical system is called maximally superintegrable when it has $2N-1$ functionally independent integrals of motion. In these  systems all trajectories are closed).
It was argued in \cite{Leonhardt06}, that the cloaking phenomenon takes place when  all trajectories of the dynamical system  defining the
refraction index are closed. In other words, dynamical system should be maximally superintegrable.
However, only the oscillator and the Coulomb problem on Euclidian spaces were considered in the mentioned paper.  At the same time,
there are their well-known generalizations to the spheres and two-sheet hyperboloids  defined by the potentials
\cite{higgs}
\be
 V_{osc}=\frac{\omega^2{\bf r}^2}{(1-\kappa{\bf r}^2)^2},\qquad V_{Coul}=-\gamma \frac{|1-\kappa {\bf r}|}{r},
\ee
as well as their further superintegrable deformations  including, in particular, the Calogero-like term \cite{CalCoul}.

Considering the energy surface $\mathcal{H}-E =0$ as a constraint \eqref{constraint}   and properly rescaling  the potential $V({\bf r})$ along with value of energy $E$, one gets
the modified profiles which can be used  for describing cloaking and perfect imaging phenomena
\begin{equation}
\mathcal{H}={(1+\kappa r^2)^2}{{\bf p}^2}+V(r), \quad \Rightarrow \quad \widetilde{n}({\bf r})=n_0\frac{\sqrt{|1-V({\bf r})|}}{|1+\kappa{\bf r}^2|}.
\end{equation}
The addition of Calogero-like term breaks spherical symmetry of the profile at the same time preserving its superintegrability.
However, in this case symmetry algebra is highly nonlinear which may cause troubles in the description of closed ray trajectories in a purely algebraic way.

Another way to find the profiles which should possess perfect imaging is to perform the simple canonical transformation $({\bf p}, {\bf r})\to (-{\bf r}, {\bf p})$. In this case the energy surface
takes a form $|{\bf r}|=n(p)$. Then expressing $p$ via $r$, we will get the new profile admitting cloaking, given by the function $\widetilde{n}_{inv}$ which is inverse to \textit{the} initial profile $n(r)$:
 $\widetilde{n}_{inv}\left( n(r)\right)=r$.
For example, it transforms the initial profile \eqref{ns}  to the one associated with the Coulomb problem
\be
(1+\kappa p^2)^2r^2=\frac{n^2_0}{\lambdabar^2_0}\quad\Rightarrow\quad   {n}_{inv}=\sqrt{\frac1{\kappa} \left(\frac{n_0}{\lambdabar_0 r}-1\right)},
\label{nCoul}\ee
\section{Inclusion of polarization}
Let us briefly discuss the inclusion of polarization.
To  this end we should add to the Lagrangian \textit{the term} ${\bf p\dot r}$, the vector-potential  of ``Berry monopole" $\mathbf{ A({p})\dot p}$ i.e. by the potential of the Dirac monopole located at the origin of \textit{the} momentum space \cite{bliokh}
\begin{equation}
\frac{\partial}{\partial \mathbf{p}}\times \mathbf{A}(p)=\frac{\mathbf{p}}{|\mathbf{p}|^3}
\label{monop}
\end{equation}
From the  viewpoint of Hamiltonian formalism this means that
 we should preserve the form of \textit{the} Hamiltonian \eqref{h0} and replace the initial Poisson brackets \eqref{pb0} by the modified ones
 \begin{equation}
  \{f,g\}=\frac{\partial f}{\partial x_\alpha}\frac{\partial g}{\partial p_\alpha}-\frac{\partial f}{\partial p_\alpha}\frac{\partial g}{\partial x_\alpha}-\frac{sp_k}{p^3}\varepsilon_{klm}\frac{\partial f}{\partial x_l}\frac{\partial g}{\partial x_m}
 \label{pbm}
 \end{equation}
  where $s$ is the spin of the photon, which is equal to one for circularly polarized photon and to zero  for linearly  polarized photon.
  The above deformation of Poisson bracket violates  the  symmetry of the Hamiltonian, and therefore, can break the closed trajectories. However the basic trajectory ($T=0, L=L_{max}$) in the limit $s\to 0$ preserves its form (see below).
  Using new definition of \textit{the} Poisson bracket Eq.(\ref{pbm}), one gets the equations of motion in the form

 \begin{eqnarray}
\dot{\bf r}=\frac{{\bf p}}{p}+\frac{s {\bf L}}{\lambdabar r p^3}\frac{\partial n}{\partial r},\quad \dot{\bf p}=\lambdabar^{-1}{\bf \nabla}n(r)\quad
\dot{\bf T}=\frac{2s\kappa {\bf L}}{\lambdabar r p^3}\frac{\partial n}{\partial r}({\bf pr}) \nonumber\\
 \dot{\bf L}=\frac{s}{\lambdabar r}\frac{\partial n}{\partial r}\left(\frac{{\bf p}({\bf pr})}{p^3}-\frac{{\bf r}}{p}\right),\quad
\dot{\bf S}=-\frac{s}{\lambdabar r}\frac{\partial n}{\partial r}\left(\frac{{\bf p}({\bf pr})}{p^3}-\frac{{\bf r}}{p}\right)
\label{eqmodexp}
\end{eqnarray}
where spin vector is determined as ${\bf S}=s{\bf p}/p$. When spin variable is taken into account in the spherical symmetrical refraction index profile $n(r)$, from Eq.(\ref{eqmodexp}) it follows that the total angular momentum ${\bf J}={\bf L}+{\bf S}$ is preserved, $\dot{\bf J}\equiv 0$. We will develop perturbation theory on $s$ exploring Eqs.(\ref{eqmodexp}). In the first-order perturbation theory on $s$, one can substitute all the terms containing $s$ by their zero-order values (values when $s=0$).
Using this approach, we can see that, in the first-order approximation, the value of $T$ on the basic trajectory is zero: $T(s)=0$.
Scalarly  multiplying \textit{$\bf J$ by $\bf r$}, for the basic trajectory, one gets ${\bf Jr}=0$. In this approximation, it follows from Eq.(\ref{trajec}) that $a_0=0,\quad R=2r_0$ and $|{\bf r}|^2=4r_0^2$. This means that basic trajectory remains a circle \textit{from} the same sphere with the center at the origin. However the plane of the circle is rotated and now is perpendicular to ${\bf J}$ and not to
${\bf L_0}$ as in $s=0$ case. The rotation angle can be found by scalarly multiplying ${\bf J}$ and ${\bf p}$: ${\bf Jp}=sp$ and therefore $\cos\theta=s/J$ and $\sin\phi=s/J$, where $\theta$ and $\phi\approx\pi/2-\theta$ are angles between $\bf J$ and $\bf p$ and $\bf J$ and $\bf L_0$, respectively (see Fig.3).
\begin{figure}
   \begin{center}
\animategraphics[autoplay,loop]{0.5}{frame}{3}{4}
\caption{Basic ray trajectories for photon with different polarizations. $\phi$ is the trajectory plane rotation angle.}
\label{fig.3}
\end{center}
\end{figure}

In the first order on $s$, $J$ can be substituted by $L_{max}=n_0r_0/\lambdabar$, \quad $sin\phi=s\lambdabar/n_0r_0\ll 1$.  Hence the actual perturbation parameter is $s\lambdabar/n_0r_0\ll 1$ therefore perturbation theory can be applied for $ s=\pm 1$ as well. The sign of rotation angle depends on the sign of $s$. So for right hand circular polarized and left hand circular polarized photons one will have different trajectories on the sphere. This is an analogue of spin Hall effect \cite{Mur04, bnature, bliokh}  in Maxwell fish eye refraction profile.



\acknowledgements
Authors are  grateful to Ashot Hakobian,  Arsen Hakhoumian, Rubik Pogossian and Khachik Nerkararian  for useful discussions and comments. This work was performed within ICTP  Affiliated Center program AF-04, and partial financial support from
Armenian Committee of Science Grants  18T-1C106 (A.N.) and 18T-1C082 (Zh.G.).

\end{document}